\documentclass[journal=nalefd,manuscript=article]{achemso}
\usepackage{graphicx,keyval}
\usepackage{amssymb, amsmath, amsfonts, wasysym}
\usepackage{float}
\usepackage{color}
\usepackage{bm}

\author{Jianfa Zhang}
\affiliation{College of Optoelectronic Science and Engineering, National University of Defense Technology, Changsha 410073, China}
\email{jfzhang85@nudt.edu.cn}
\author{Zhihong Zhu}
\affiliation{College of Optoelectronic Science and Engineering, National University of Defense Technology, Changsha 410073, China}
\alsoaffiliation{State Key Laboratory of High Performance Computing, National University of Defense Technology, Changsha 410073, China}
\author{Wei Liu}
\affiliation{College of Optoelectronic Science and Engineering, National University of Defense Technology, Changsha 410073, China}
\author{Xiaodong Yuan}
\affiliation{College of Optoelectronic Science and Engineering, National University of Defense Technology, Changsha 410073, China}
\author{Shiqiao Qin}
\affiliation{College of Optoelectronic Science and Engineering, National University of Defense Technology, Changsha 410073, China}
\alsoaffiliation{State Key Laboratory of High Performance Computing, National University of Defense Technology, Changsha 410073, China}

\title[Graphene light trapping] {Graphene plasmonics for light trapping and absorption engineering}

\begin{document}

%%%%%%%%%%%%%%%%%%%%%%%%%%%%%%%%%%%%%%%%%%%%%%%%%%%%%%%%%%%%%%%%%%%%%
%% The "tocentry" environment can be used to create an entry for the
%% graphical table of contents. It is given here as some journals
%% require that it is printed as part of the abstract page. It will
%% be automatically moved as appropriate.
%%%%%%%%%%%%%%%%%%%%%%%%%%%%%%%%%%%%%%%%%%%%%%%%%%%%%%%%%%%%%%%%%%%%%
%\begin{tocentry}
%The surrounding frame is 9\,cm by 3.5\,cm.
%\includegraphics[width=90mm]{FigTOC.eps}
%\end{tocentry}

%%%%%%%%%%%%%%%%%%%%%%%%%%%%%%%%%%%%%%%%%%%%%%%%%%%%%%%%%%%%%%%%%%%%%
%% The abstract environment will automatically gobble the contents
%% if an abstract is not used by the target journal.
%%%%%%%%%%%%%%%%%%%%%%%%%%%%%%%%%%%%%%%%%%%%%%%%%%%%%%%%%%%%%%%%%%%%%
\begin{abstract}
Plasmonics can be used to improve absorption in optoelectronic devices and has been intensively studied for solar cells and photodetectors. Graphene has recently emerged as a powerful plasmonic material. It shows significantly less losses compared to traditional plasmonic materials such as gold and silver and its plasmons can be tuned by changing the Fermi energy with chemical or electrical doping. Here we propose the usage of graphene plasmonics for light trapping in optoelectronic devices and show that the excitation of localized plasmons in doped, nanostructured graphene can enhance optical absorption in its surrounding media including both bulky and two-dimensional materials by tens of times, which may lead to a new generation of highly efficient, spectrally selective photodetectors in mid-infrared and THz ranges. The proposed concept could even revolutionize the field of plasmonic solar cells if graphene plasmons in the visible and near-infrared are realized.
\end{abstract}


%%%%%%%%%%%%%%%%%%%%%%%%%%%%%%%%%%%%%%%%%%%%%%%%%%%%%%%%%%%%%%%%%%%%%
%% Start the main part of the manuscript here.
%%%%%%%%%%%%%%%%%%%%%%%%%%%%%%%%%%%%%%%%%%%%%%%%%%%%%%%%%%%%%%%%%%%%%
%\section{Introduction}
Graphene, a single layer of carbon atoms arranged in plane with a honey comb lattice, shows promising potentials in optics and optoelectronics~\cite{bonaccorso2010graphene}. Among its many novel properties, the collective electronic excitations, known as graphene plasmons, is one of the most attractive ones~\cite{grigorenko2012graphene,bao2012graphene,low2014graphene,garcia2014graphene}. Graphene plasmons have been demonstrated through spectral characteristics of light scattering by graphene nanoribbons/disks in the infrared and THz ranges~\cite{ju2011graphene,yan2012tunable} and observed directly with scanning near-field optical microscopy~\cite{chen2012optical,fei2012gate}. Even though the interaction between light and graphene is supposed to be quite weak and a monolayer graphene shows an optical absorption of only about $2.3\%$ in the visible and near infrared range, the excitation of graphene plasmons totally changes this picture. The excitation of propagating surface plasmons in graphene makes it possible to guide light with deep subwavelength mode profiles~\cite{vakil2011transformation}. Meanwhile, doped and patterned graphene can support localized plasmonic resonances, leading to efficient confinement of light and strong enhancement of local fields~\cite{brar2013highly,jang2014tunable}. Thus, graphene plasmonics provide an effective route to enhance light-graphene interactions~\cite{koppens2011graphene}. The exploration of graphene plasmonics has lead to the proposition and demonstration of a variety of functionalities in mid-infrared and THz ranges such as graphene waveguides~\cite{vakil2011transformation}, photodetectors~\cite{freitag2013photocurrent}, tunable metamaterials~\cite{tassin2012comparison,papasimakis2013magnetic}, filters and polarizers~\cite{Zhu2014,zhu2014electrically}, and others. Essentially, graphene is retelling the story of plasmonics~\cite{de2013graphene}.
%applicaitons: sensing, photovoltaics, nonlinear optics

One of the most important applications for plasmonics is to enhance the optical absorption in optoelectronic devices~\cite{green2012harnessing}. In solar cells and photodetectors, the material extinction must be high enough to allow efficient light harvesting and photocarrier generation. On the other hand, there is a strong desire to reduce the thickness or volume of semiconductors in these devices in order to decrease the consumption of materials, reduce the material deposition requirements, and/or improve the performances (e.g., increase the collection of minority carriers, detectivity or time responses). Moreover, two-dimensional materials have recently emerged as promising candidates for optoelectronic applications~\cite{lopez2014light,xia2014two,koppens2014photodetectors}. Even though such materials have quite high quantum efficiencies for light-matter interactions, their absorption is quite weak in absolute terms. As a result, absorption engineering is of great significance. Plasmonic have been demonstrated to be one of the most effective routes for light trapping and absorption enhancement~\cite{atwater2010plasmonics,konstantatos2010nanostructured}. A range of different plasmonic structures such as metallic nanoparticles~\cite{catchpole2008plasmonic,nakayama2008plasmonic, chen2012broadband,chen2013exceeding}, gratings~\cite{munday2010large, min2010enhancement} , antennas~\cite{knight2011photodetection, fang2012graphene} and others~\cite{schuller2010plasmonics} have been used to improve the performance of solar cells and photodetectors including those built with two-dimensional materials.

Graphene exhibits remarkably less losses compared to traditional plasmonic materials such as noble metals (e.g., gold and silver) and is very promising for light trapping in optoelectronic devices~\cite{garcia2014graphene}. Moreover, the ability of being electrically tunable makes it possible to realize active spectral selectivity~\cite{chen2012optical,fei2012gate}, which is a favorable property for photodetection~\cite{laux2008plasmonic}. Enhancement of optical absorption by graphene plasmonics have been intensively studied. Perfect absorption have been theoretically predicted in nanostructured graphene backed by a mirror or in a free standing structured graphene film under the illumination of two coherent beams~\cite{thongrattanasiri2012complete,zhang2014coherent}. Recently, more than one order of absorption enhancement in arrays of doped graphene disks have been experimentally demonstrated~\cite{fang2013gated,fang2013active}. However, most of previous studies have been focused on engineering the absorption in the graphene itself and the potential of using the low-loss graphene plasmons to enhance the absorption of other light-absorbing materials have just not been fully realized. In the letter, we propose a type of hybrid optoelectronic devices based on the integration of graphene plasmonic structures with bulky semiconductors or two-dimensional materials. We show numerically that the excitation of localized plasmons in doped, nanostructured graphene provides a very effective way for light trapping and can significantly enhance the absorption in surrounding light-absorbing materials, which may lead to a new generation of highly efficient, spectrally selective and tunable photodetectors~\cite{xu2010plasmonic,liu2011plasmon,sobhani2013narrowband,freitag2013photocurrent}.

% The plasmon induced charge density for a single nanodisk.
Figure 1a and 1b shows the schematic illustration and geometric parameters of our proposed device. An array of doped periodical graphene nanodisks is integrated with a layer of light-absorbing materials. There is an insulator layer between them. The thicknesses of these two layers are $t$ and $s$, respectively. The substrate is assumed to be semi-infinite. The period of the graphene arrays is $P=400$~nm and the diameter of graphene disks is $D=240$~nm. Both the substrate and the insulator layer are assumed to be lossless with a dielectric constant of 1.96. The dielectric constant (real part) of the light-absorbing materials is $\epsilon'=10.9$ and the losses are introduced through the imaginary part $\epsilon''$ of the dielectric constant, which is related to the attenuation coefficient $\alpha = -(2\pi/\lambda)Im(\sqrt{\epsilon' + i\epsilon''})$.
% Absorbed photons are converted into electron-hole pairs %$\epsilon"=-(\lambda/\pi)\sqrt{\epsilon'+(\lambda\alpha/2\pi)^{2}}\alpha

The numerical simulations are conducted using a fully three-dimensional finite element technique (in Comsol MultiPhysics). In the simulation, the graphene is modelled as a conductive surface~\cite{thongrattanasiri2012complete, vakil2011transformation, yao2013broad}. The sheet optical conductivity of graphene can be derived within the random-phase approximation (RPA) in the local limit~\cite{falkovsky2007optical,falkovsky2007space}
%The in-plane permittivity of graphene can be calculated from the graphene sheet optical conductivity.
%--------------------------------------------------------------
\begin{equation}
\begin{split}
\sigma_{\omega}=  & \frac{2e^{2} k_{B} T}{\pi \hbar^{2}} \frac{i}{\omega+i\tau^{-1}} ln[2\cosh(\frac{E_{F}}{2k_{B} T})] + \\
&\frac{e^{2}}{4\hbar}[\frac{1}{2}+\frac{1}{\pi} arctan(\frac{\hbar \omega-2E_{F}}{2k_{B} T})-\\
& \frac{i}{2\pi}\ln \frac{(\hbar \omega+2E_{F})^{2}}{(\hbar \omega-2E_{F})^{2}+4(k_{B}T)^{2}}]
\end{split}
\end{equation}
%--------------------------------------------------------------
Here $k_{B}$ is the Boltzmann constant, $T$ is the temperature, $\omega$ is the frequency of light, $\tau$ is the carrier relaxation lifetime, and $E_{F}$ is the Fermi energy. The first term in Eq.~(1) corresponds to intra-band transitions and the second term is attributed to inter-band transitions. We restrict our calculations to photon energies below 0.2~eV and to the Fermi level $E_{F}\gg 2k_{B} T$. Moreover, the Fermi level is increased above half of the photon energy so the contribution of inter-band transition can be avoided due to Pauli blocking. Equation~(1) reduces to the Drude model if we neglect both inter-band transitions and the effect of temperature ($T=0$)~\cite{hanson2008quasi,jablan2009plasmonics}
%--------------------------------------------------------------
\begin{equation}
\sigma_{\omega}=\frac{e^{2} E_{F}}{\pi \hbar^{2}} \frac{i}{\omega+i\tau^{-1}}
\end{equation}
%--------------------------------------------------------------
where $E_{F}$ depends on the concentration of charged doping and $\tau=\mu E_{F}/(ev_{F}^{2})$, where $v_{F}\approx 1\times 10^{6}~m/s$ is the Fermi velocity and $\mu$ is the dc mobility. Here we use a moderate measured mobility $\mu=10000~cm^{2}\cdot V^{-1} \cdot s^{-1}$~\cite{novoselov2004electric}. At first, we assume the Fermi energy of graphene to be $E_{F}=0.6$~eV which corresponds to a doping density of about $2.6 \times 10^{13}~cm^{-2}$ and may be realized by electrostatic or chemical doping~\cite{novoselov2004electric, fang2013gated}.

Figure 1c shows the numerically simulated spectra under the illumination of a plane wave at normal incidence. Here the thickness of the insulator layer is $s=20~nm$. The light-absorbing layer is $t=100~nm$ thick with an absorption coefficient $\alpha=-0.1~\mu m^{-1} $ corresponding to a small absorption of only about $2\%$ in impedance matched media. There is a resonance at around $15.4~\mu m$ in the spectra with strong light extinction. The total absorption is $A=36.9\%$ while the absorbance by the absorptive layer reaches $A'=25.4\%$, representing an enhancement of about 12.5 times. As shown in Figure 1d, this resonance is attributed to the excitation of a dipolar plasmonic mode in the doped graphene nanodisks. The oscillation of localized surface plasmons leads to the light trapping and local field enhancement around the graphene nanodisks and serves to the enhancement of absorption in the absorptive layer nearby (see Figure 1e). As the graphene nanodisk is isotropic, the optical response here is independent of polarization at normal incidence~\cite{hopkins2013optically}.

Even though graphene is much less lossy compared to traditional plasmonic metals such as gold and silver, part of light is inevitably absorbed in the graphene nanodisks. The competition of absorption between the graphene and the absorptive layer underneath depends significantly on the absorptive coefficient of the later. Figure 2 shows absorption spectra (total absorption $A$ and absorption in the underlying absorptive layer $A'$) with different absorption coefficients. When the absorption coefficient is $\alpha=-0.05~\mu m^{-1}$ (Figure 2a) corresponding to an absorption of only $1\%$ in an impedance matched medium, $20.5\%$ of the incident light will be absorbed by the thin layer of absorptive medium, which accounts for about half of the total absorption ($\sim 40.9\%$). Note that even though the absolute absorption in the layer is slightly lower compared to the ($\sim 25.4\%$) absorption for $\alpha=-0.1~\mu m^{-1}$, the enhancement factor of absorption here reaches 20 and it is 1.6 times higher due to the decreased absorption losses of the total system and the increased quality factor of resonance. As the absorption coefficient of the layer increases to $\alpha=-0.2~\mu m^{-1}$, the absorption in the absorptive layer increases to $\sim27.5\%$ while the total absorption is $\sim 33.6\%$ at the resonance (Figure 2b). A majority of the light is now absorbed by the layer underneath. The absorptive layer will now have an absorption of about $\sim 3.9\%$ in an impedance matched medium, so the absorption enhancement factor is about 7. At the off-resonance wavelengths, the absorption of the light-absorbing layer with graphene is almost the same as that without graphene, which is about two times of that in an impedance matched medium due to the light trapping by interference effects. This can be an advantage of graphene compared to metallic nanostructures for light trapping as the scattering of the later can sometimes shadow the benefits of absorption enhancement at off-resonance wavelengths if they are located at the front side.

%Here, the effective absorption happens in the absorptive layer where absorbed photons is supposed to contribute to the generation of electron-hole pairs.
%So the absorptive layer should be located near the graphene nanodisk array in order to have enough overlap with the local electromagnetic field.
Graphene plasmons have very small spatial extensions compared with the wavelength of light in the vacuum and the localized graphene plasmons here are highly confined to regions around the nanodisks. Figure 3 shows the absorption spectra with different separations between the graphene nanodisks and the light-absorbing layer. Here the absorption coefficient is set to be $\alpha=-0.1~\mu m^{-1}$. As the thickness of the insulator layer is $s=10~nm$, the total absorption and absorption in the absorptive layer underneath are $\sim 33.4\%$ and $\sim 26.5\%$, respectively. As the separation increases,the resonance becomes sharper because the surrounding media of graphene nanodisks become less lossy and the total absorption increases until it reaches the absorption limit of about $42\%$ (i.e., $1/(1+1.4)$)~\cite{thongrattanasiri2012complete}. However, the light extinction by the absorptive layer decreases due to a reduced proportion of confined field in the layer. With a separation of $s=50~nm$, the total absorption goes up to $\sim 40.6\%$ but the light extinction by the absorptive layer reduces to $\sim 14.7\%$. At the same time, the resonance wavelength blue-shifts from $17.5~\mu m$ to $13.3~\mu m$ as the separation increases from $s=10~nm$ to $s=50~nm$. This is because the absorptive layer has a much higher dielectric constant (real part) compared to the insulator layer and air and it has a strong influence on the effective refractive index of the medium surrounding graphene (and thus the effective wavelength of light).

Figure 4 shows the spectra of absorption in the absorptive layer with different Fermi energies of graphene. When the Fermi energy $E_{F}=0.4~eV$, the resonance is at $18.87~\mu m$ and the resonant absorption is $\sim 18.9\%$. As the Fermi energy increases to $E_{F} = 0.8~eV$, the resonance blue shifts to $13.32~\mu m$ and the resonant absorption goes up to $\sim 29.7\%$. Furthermore, with a Fermi energy of $E_{F}=1.2~eV$, the resonance happens at around $10.88~\mu m$ and the resonant absorption reaches $\sim 31.8\%$. With the increase of Fermi energy, the wavelength of graphene plasmons becomes longer which is responsible for the blueshift of resonances~\cite{chen2012optical,fei2012gate}. At the same time, the conductivity of graphene increases and the graphene plasmon becomes less lossy as the Fermi energy increases. So a larger proportion of light is absorbed in the absorptive layer with higher Fermi energies.

We have also studied the angle dependence of absorption and the results are shown in Figure 5. Here the Fermi energy, absorption coefficient and geometric parameters are the same as in Figure 1 ($E_{F}=0.6~eV$, $\alpha=-0.1~\mu m^{-1}$, $s=20~nm$). The incidence-angle and polarization dependence of the absorption is quite weak for incident angle below 50 degrees. For even larger incident angles, the absorption becomes more dependent on the incidence-angle and polarization but the resonance wavelength keeps nearly the same. These results agree well with previous studies~\cite{thongrattanasiri2012complete}. The property of nearly omnidirectional absorption is beneficial for practical applications.

In the past few year, two-dimensional atomic crystals and their heterostructures have emerged as promising materials for photodetection and other optoelectronic applications~\cite{britnell2013strong,liu2014graphene,eda2013two,lopez2014light,koppens2014photodetectors}. Besides graphene, a variety of other two-dimensional materials have been studied, such as molybdenum disulphide (MoS2)~\cite{lopez2013ultrasensitive} and tungsten diselenide (WSe2)~\cite{ross2014electrically,baugher2014optoelectronic, pospischil2014solar}. These materials have very high quantum efficiencies for light-matter interactions, but their absorption is quite weak in absolute terms. So improving their interactions with light through plasmonic trapping become even more desirable. Now we replace the bulky light-absorbing layer in Figure 1 with a two-dimensional absorptive film. The absorption of the two-dimensional film can be described by its conductivity. As we know, the $2.3\%$ universal absorbance of monolayer graphene corresponds to an optical conductance of $ G_{0} = e^{2}/(4 \hbar) \approx 6.08 \times 10^{-5} ~\Omega^{-1}$~\cite{nair2008fine,stauber2008optical}. Here the conductance of the two-dimensional film is assumed to be $G_{0} = 2.65 \times 10^{-5}~\Omega^{-1}$ which corresponds to an absorption of $1\%$ in the air and is comparable to that of graphene at $\lambda = 15~\mu m$ for $E_{F} = 0.05~eV$ and $T = 150~K$ according to Eq.~1. Figure 6 shows the spectra of absorption in the underneath two-dimensional film and its enhancement factor when the graphene nanodisks are at different Fermi energies ranging from $E_{F}=0.4 ~eV$ to $1.2~eV$. When the Fermi energy $E_{F}=0.4~eV$, the resonance happens at $15.5~\mu m$ and the resonant absorption of the two-dimensional film is $\sim 17.1\%$ with an enhancement factor of 17. As the Fermi energy increases to $E_{F} = 1.2~eV$, the resonance blue shifts to $8.9~\mu m$ and the resonant absorption goes up to $\sim 36.2\%$ with an enhancement exceeding 36 times. Meanwhile, the absorption in the two-dimensional film without graphene nanodisks (data not shown in the figure) is about $0.7\%$ which is lower than that of a free standing film in the air~\cite{stauber2008optical}.

In our simulations, the size of the graphene nanodisks is fixed. Indeed, spectral tuning of absorption can also be realized through geometric variations of graphene nanodisks. Moreover, polarization dependent absorption and broadband absorption are possible with suitable designs of graphene nanostructures~\cite{zhihong2015broadband}. Here the total thickness of the graphene film combined with the insulator and absorptive layer is much smaller than the wavelength of light, so the combination of them can be regarded as a thin film with an asymmetric dielectric environment (air with the refractive index of 1 and the substrate with the refractive index of 1.4). According to previous study, the maximum absorption of such a film is limited to about $42\%$ (i.e., $1/(1+1.4)$)~\cite{thongrattanasiri2012complete}. Perfect light absorption in the whole structure and an further enhancement of absorption in the underlying absorptive layer can be achieved by adding a metallic reflecting mirror. Furthermore, one can also replace the graphene nanodisk array with a graphene film patterned with nanoholes, which may be more convenient for electrical tuning of the Fermi energy.

In summary, we show numerically that graphene plasmonics provides a very effective way for light trapping and absorption engineering in optoelectronic devices. The excitation of localized plasmons in doped, nanostructured graphene can significantly enhance the light-matter interactions and lead to strong absorption. As graphene plasmons are much less lossy compared to plasmons of coinage metals, a large proportion of the light can be absorbed by the absorptive material (e.g., semiconductors or two-dimensional materials) surrounding the graphene nanostructure with an enhancement up to tens of times at the resonance. Moreover, the electrical tunability of graphene plasmons makes it possible for active spectral selectivity. Even though we have just discussed the exploration of localized graphene plasmons, propagating graphene surface plasmons may also be explored for light trapping and absorption enhancement~\cite{yin2014plasmonic}. Graphene plasmons have so far been observed at mid-infrared and longer wavelengths, so the proposed concept is particularly promising for potential applications in infrared and THz photodetection. However, it has been theoretically suggested that they can be extended toward the visible and near-infrared range by several strategies including a reduction in the size of the graphene structures and an increase in the level of doping~\cite{garcia2014graphene}. If such theoretical prediction is realized, it may lead to a revolution in photovoltaics by replacing noble metals such as gold and silver with graphene for light harvesting in plasmonic solar cells.

%\section{author information}
%\textbf{Corresponding Author}
%jianfa zhang, E-mail: jfzhang@nudt.edu.cn.
%\textbf{Competing financial interests}
%\section{Additional information}
%The authors declare no competing financial interests.

%%%%%%%%%%%%%%%%%%%%%%%%%%%%%%%%%%%%%%%%%%%%%%%%%%%%%%%%%%%%%%%%%%%%%
%% The "Acknowledgement" section can be given in all manuscript
%% classes.  This should be given within the "acknowledgement"
%% environment, which will make the correct section or running title.
%%%%%%%%%%%%%%%%%%%%%%%%%%%%%%%%%%%%%%%%%%%%%%%%%%%%%%%%%%%%%%%%%%%%%
\begin{acknowledgement}
The work is supported by National Natural Science Foundation of China (grant nos. 11304389, 61177051, 61205087 and 11404403) and the National Basic Research Program of China (973 Program, grant no. 2012CB933501).
\end{acknowledgement}

%%%%%%%%%%%%%%%%%%%%%%%%%%%%%%%%%%%%%%%%%%%%%%%%%%%%%%%%%%%%%%%%%%%%%
%% The same is true for Supporting Information, which should use the
%% suppinfo environment.
%%%%%%%%%%%%%%%%%%%%%%%%%%%%%%%%%%%%%%%%%%%%%%%%%%%%%%%%%%%%%%%%%%%%%
%\begin{suppinfo}
%Spectral tuning of absorption by changing the diameter of graphene nanodisks. Perfect light absorption in the whole structure and an further enhancement of absorption in the underlying absorptive layer with a back mirror. Light trapping and enhancement of absorption by a doped graphene sheet with an array of periodical nano-holes.
%\end{suppinfo}

%%%%%%%%%%%%%%%%%%%%%%%%%%%%%%%%%%%%%%%%%%%%%%%%%%%%%%%%%%%%%%%%%%%%%
%% The appropriate \bibliography command should be placed here.
%% Notice that the class file automatically sets \bibliographystyle
%% and also names the section correctly.
%%%%%%%%%%%%%%%%%%%%%%%%%%%%%%%%%%%%%%%%%%%%%%%%%%%%%%%%%%%%%%%%%%%%%
%\bibliography{Graphene}
\providecommand{\latin}[1]{#1}
\providecommand*\mcitethebibliography{\thebibliography}
\csname @ifundefined\endcsname{endmcitethebibliography}
  {\let\endmcitethebibliography\endthebibliography}{}

%\section{Results and discussion}
\begin{figure*}
\centering
\includegraphics[width=170mm]{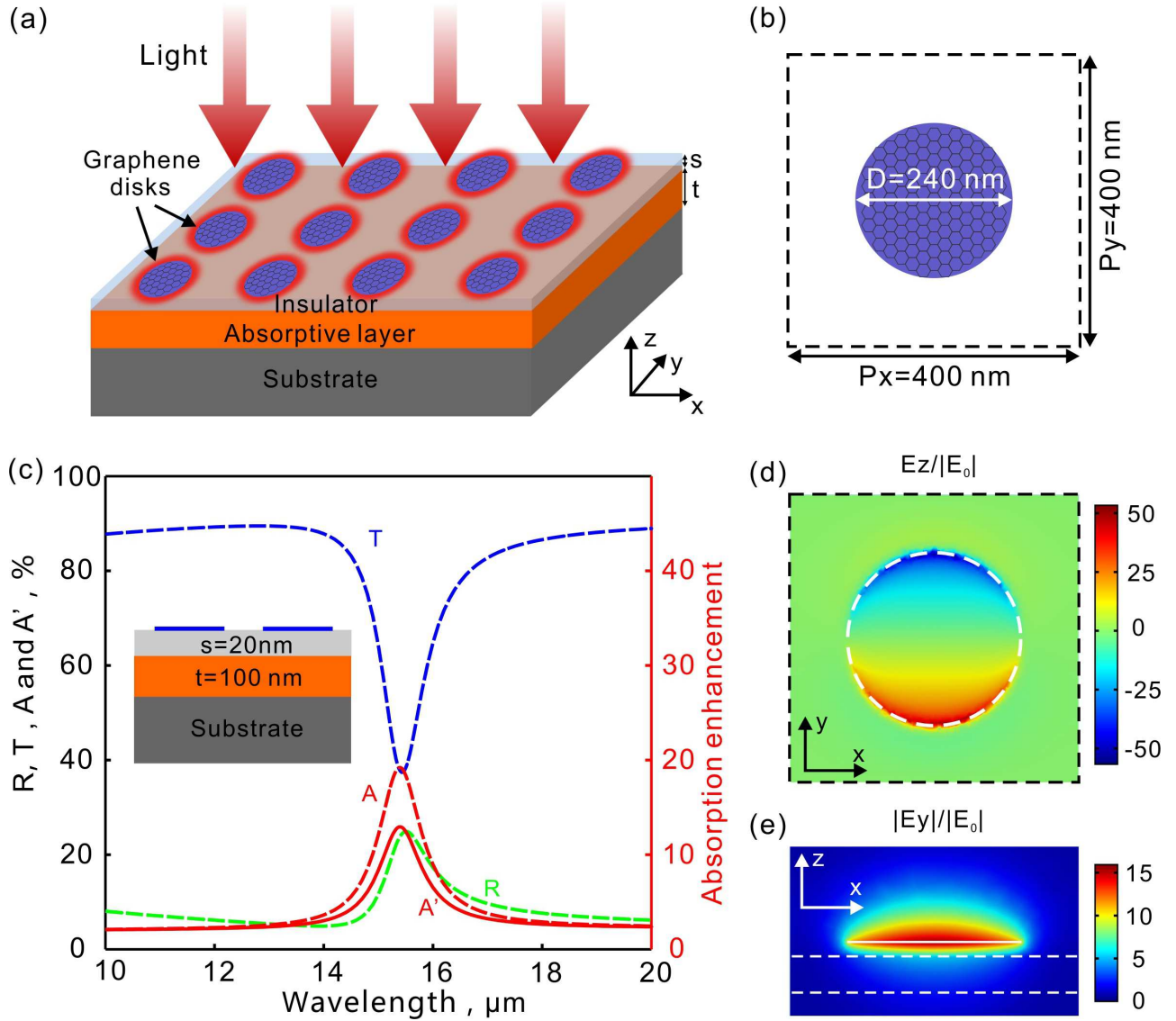}
\caption{Improving light harvesting with graphene plasmonics. (a) Scheme of the proposed devices. From the top to the bottom of the structure are an array of graphene nanodisks, an insulator layer with a thickness of $s$, the absorptive layer with a thickness of $t$ and a semi-infinite substrate, respectively. Incident light excites localized plasmons in the doped graphene nanodisks, which trap light in the near-field and enhance optical absorption in the light-absorbing layer underneath. (b) A unit cell of the graphene nanodisk array. The period is $P=Px=Py=400~nm$ and the diameter of the graphene disk is $D=240~nm$. (c) Simulated spectra of reflection ($R$), transmission($T$), absorption ($A$) as well as absorption in the absorptive layer ($A'$) with the Fermi energy at $E_{F}=0.6~eV$. The enhancement of absorption in the absorptive layer is also shown. (d) Electric field in z-direction. The field is normalized to the field amplitude of the incident light ($E_{0}$) and plotted at the x-y plane that is 5~nm above the graphene disks. (e) Normalized electric field in y-direction at the x-z plane bisecting the graphene disks.}
\end{figure*}

\begin{figure}
\centering
\includegraphics[width=82.5mm]{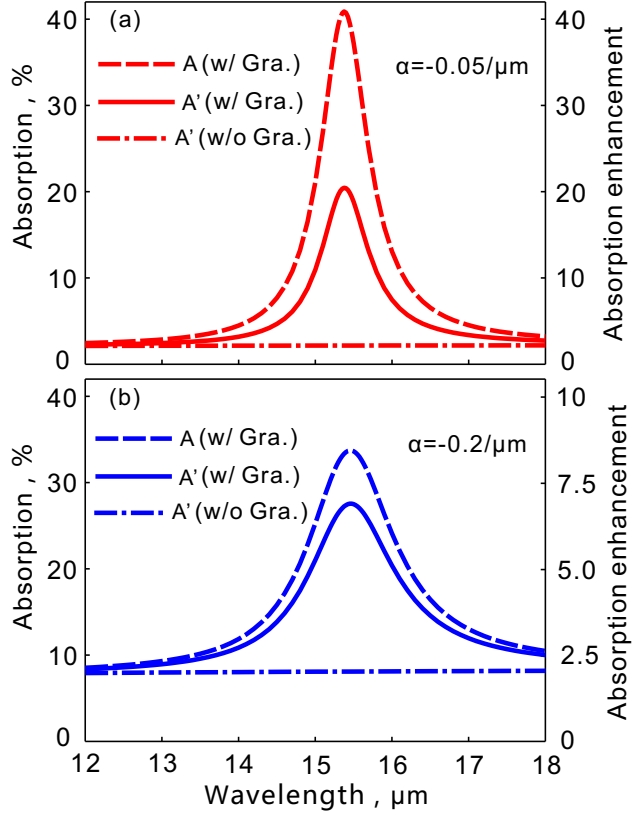}
\caption{ Simulated spectra of absorption (total absorption $A$ and absorption in the underlying absorptive layer $A'$) with different absorption coefficients $\alpha=-0.05~\mu m^{-1} $ and $-0.2~\mu m^{-1} $. The enhancement factor of absorption in the absorptive layer is also shown compared to that in an impedance matched medium. As a reference, the absorption in the light-absorbing layer without graphene is also shown (the flat dot-dashed line).}
\end{figure}

\begin{figure}
\centering
\includegraphics[width=82.5mm]{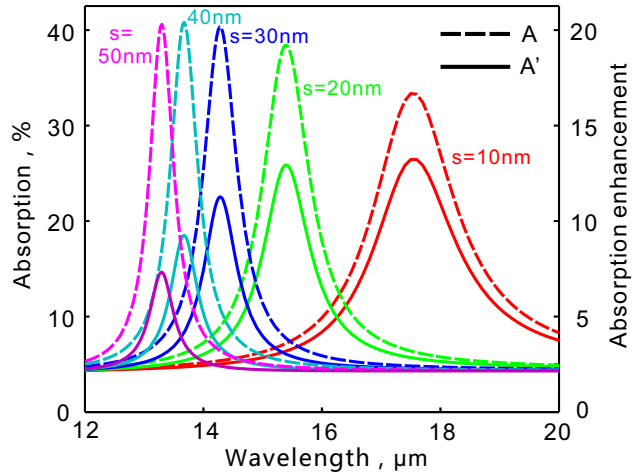}
\caption{ Simulated spectra of total absorption and absorption in the underlying absorptive layer with different separations between the graphene nanodisk array and the absorptive layer. The separation ranges from $s=10$ to $50~nm$. The enhancement factor of absorption in the absorptive layer is also shown compared to that in an impedance matched medium.}
\end{figure}

\begin{figure}
\centering
\includegraphics[width=82.5mm]{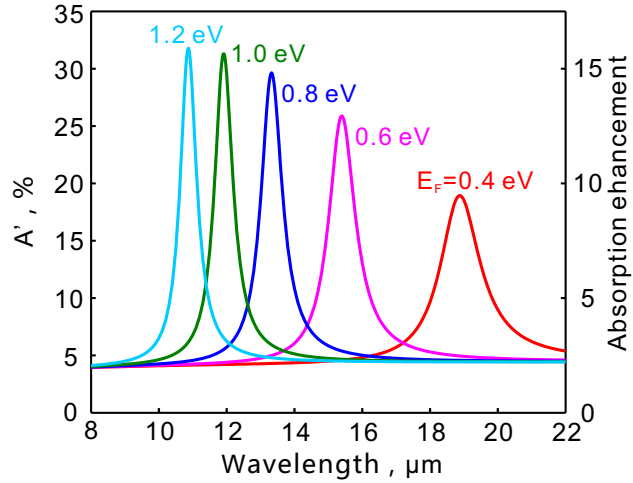}
\caption{Spectral tunability of absorption with variations of Fermi energy. The absorption in the absorptive layer with the Fermi energy of graphene ranging from 0.4 to 1.2~eV are shown along with the enhancement factor compared to that in an impedance matched medium. The separation is $s=20$~nm.}
\end{figure}

\begin{figure}
\centering
\includegraphics[width=160mm]{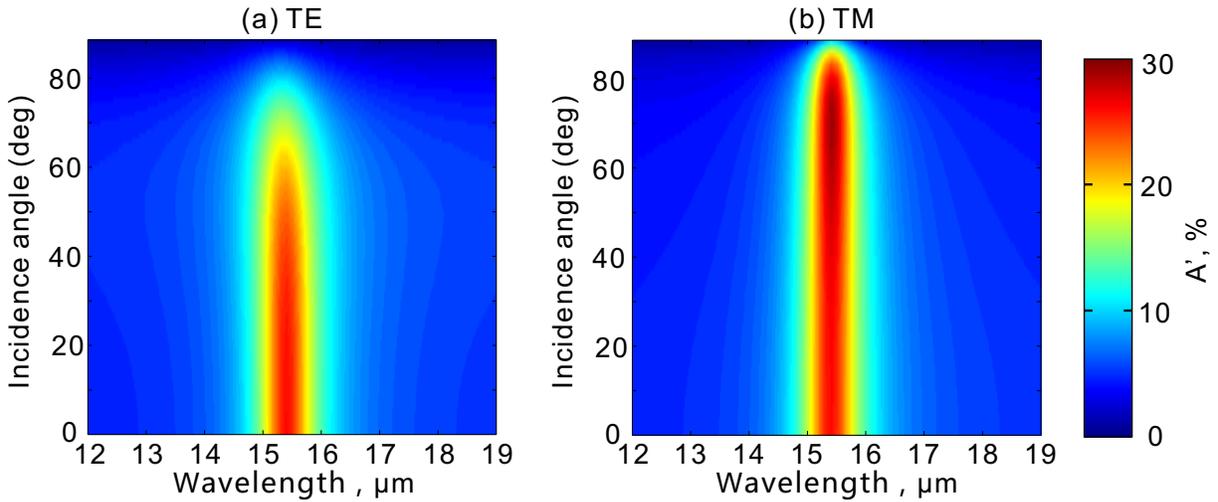}
\caption{Angular dispersions of the resonant absorption in the absorptive layer for (a) s-polarized (TE) and (b) p-polarized (TM) light. Here the absorption coefficient of the absorptive layer is $\alpha=-0.1~\mu m^{-1}$. The Fermi energy of graphene is $E_{F}=0.6~eV$ and the separation is $s=20~nm$. }
\end{figure}

\begin{figure}
\centering
\includegraphics[width=82.5mm]{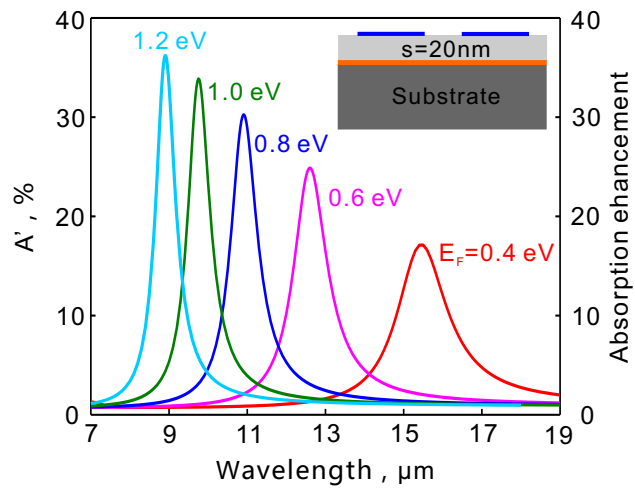}
\caption{Absorption enhancement in two-dimensional materials through light trapping with graphene plasmonics. The absorption in the unstructured, light-absorbing two-dimensional material with the Fermi energy of graphene ranging from 0.4 to 1.2~eV are shown along with enhancement factor compared to absorption of the free-standing two-dimensional material in air. The separation is $s=20~nm$.}
\end{figure}

\end{document}